\documentclass[aps,twocolumn]{revtex4}

\usepackage{graphics}
\usepackage{graphicx}
\usepackage{bm}
\usepackage{latexsym}
\usepackage{color, soul}
\usepackage{mathrsfs}
\usepackage{braket}
\usepackage{enumerate}
\usepackage[usenames,dvipsnames]{xcolor}
\usepackage[colorlinks=true,citecolor=blue,linkcolor=blue,urlcolor=blue,backref=true]{hyperref}
\usepackage{float}
\usepackage[raggedright,tight]{subfigure}  %side-by-side figures
\usepackage{amsmath}
\usepackage{amsfonts}
\usepackage{braket}
\usepackage[normalem]{ulem}
\usepackage{cancel}
\usepackage{mathtools}
\usepackage{appendix}

\begin{document}

\title{Arbitrary gauge quantisation of light-matter theories with time-dependent constraints}

\author{Adam Stokes$^1$}
\author{Ahsan Nazir$^2$}

\affiliation{$^1$School of Mathematics, Statistics, and Physics, Newcastle University, Newcastle upon Tyne NE1 7RU, United Kingdom}
\affiliation{$^2$Department of Physics and Astronomy, University of Manchester, Oxford Road, Manchester M13 9PL, United Kingdom}

\begin{abstract}
We provide a general framework for the quantisation of light-matter theories with time-dependent holonomic constraints. Unless time dependence is present from the outset at the Lagrangian level, different gauges generally produce non-equivalent canonical theories. The irrotational gauge is defined as that which also yields a correct theory when time dependence is introduced at the Hamiltonian level. Our framework unifies examples of such gauges found in existing literature. In particular, we show that for describing time-dependent light-matter interactions the Coulomb gauge is not generally irrotational, so it does not enjoy any special status.
\end{abstract}

\maketitle

\section{Introduction}

Controlling microscopic light-matter interactions has an almost unbounded range of applications. For example, the active control of spacetime metamaterials for microwave engineering, photonics, and even acoustics, is rapidly progressing \cite{caloz_spacetime_2020,caloz_spacetime_2020-1,pacheco-pena_merging_2023,galiffi_photonics_2022}, while quantum descriptions of time-varying media have also revealed novel phenomena \cite{horsley_quantum_2023,pendry_qed_2024,horsley_macroscopic_2025}. The external control of superconducting devices allows state control \cite{liu_optical_2005} and stabilisation \cite{huang_universal_2018,lu_universal_2017}, quantum switches \cite{mariantoni_two-resonator_2008}, entanglement harvesting \cite{armata_harvesting_2017}, fast quantum gates \cite{beaudoin_first-order_2012,strand_first-order_2013,romero_ultrafast_2012,Groszkowski_tunable_2011}, and improved qubit connectivity \cite{naik_random_2017,didier_analytical_2018}. Subcycle interaction switching with quantum wells \cite{gunter_sub-cycle_2009} and ultrastrong switching with superconducting circuits \cite{peropadre_switchable_2010} were achieved some time ago, and more recently using an electron gas \cite{halbhuber_non-adiabatic_2020}. Controlled ultrastrong interactions within electromagnetic cavities containing atoms and molecules \cite{flick_lightmatter_2019,keyl_controlling_2014,kowalewski_non-adiabatic_2016,leroux_enhancing_2018}, and solid-state emitters \cite{falci_ultrastrong_2019} also possesses numerous applications, including chemical reaction control \cite{garcia-vidal_manipulating_2021,hertzog_strong_2019}.

Modeling direct control requires the introduction of prescribed time dependence, which is also important for understanding loss and decoherence \cite{garziano_switching_2013,wellstood_lowfrequency_1987,kumar_origin_2016,yoshihara_decoherence_2006}, as well as fundamental vacuum phenomena, such as the Unruh effect \cite{unruh_what_1984,lopp_quantum_2021,passante_radiative_1998} and the dynamical Casimir effect \cite{rossatto_entangling_2016,dodonov_fifty_2020}. The simplest method to achieve this is to swap out certain model parameters for time-dependent functions, but whether such a procedure is truly justified will often be unclear. In particular, it will generally result in non-equivalent theories when implemented in different gauges \cite{stokes_implications_2022}. A number of recent articles have addressed the question of if and how the resulting descriptions may or may not be valid \cite{stokes_implications_2022,stokes_ultrastrong_2021,settineri_gauge_2021,gustin_gauge-invariant_2023,di_stefano_resolution_2019}.

A more refined approach, pursued here, is to prescribe physically motivated time-dependent constraints. We begin in Sec.~\ref{general} with a general presentation of the ways in which time dependence can be included in both Lagrangian and Hamiltonian (canonical) theories with holonomic constraints. We demonstrate that, in general, only if it is present within the constraints from the outset of the Lagrangian description, does a {\em unique} canonical description follow. This description in turn enables one to determine if the more naive approach of introducing time dependence phenomenologically at a later stage, is valid, and if it is, which gauge provides the valid naive theory. The name given to this gauge is {\em irrotational} \cite{you_circuit_2019}. In Sec.~\ref{examples} we use the framework of Sec.~\ref{general} to consolidate disparate examples of the irrotational gauge found within existing literature \cite{stokes_ultrastrong_2021,you_circuit_2019}. The previously debated status of the Coulomb gauge for describing time-dependent light-matter interactions is determined conclusively by our demonstration that it does not generally coincide with the irrotational gauge. In Sec.~\ref{discuss}, we explain how our results relate to those within existing literature. %explain precisely how and why the treatments of time dependence in Refs.~\cite{settineri_gauge_2021,gustin_gauge-invariant_2023,di_stefano_resolution_2019} that lead to the contrary conclusion, are not generally valid, including a discussion of examples. We conclude briefly in Sec.~\ref{conc}.

\section{Time-dependent constraints}\label{general}

\subsection{Elimination of coordinates using constraints}

For simplicity, we restrict our attention to systems described by $N$ discrete coordinates and velocities $\{x_n, {\dot x}_n\}$ with $n=0,\, 1,..., N-1$, noting that the extension to include continuously labelled coordinates (fields), does not result in different conclusions. Without loss of generality we may begin with a Lagrangian, $L\{x_n,{\dot x}_n\}$, that does not depend on time $t$. It is however, assumed to be accompanied by a finite number of generally time-dependent holonomic constraints, $f_\mu^t\{x_n\} = 0$, where $\mu =0,\,1,...,M-1$ with $M<N$. 

The constraints allow $M$ of the coordinates $x_n$ to be eliminated in favour of $N-M$ unconstrained coordinates $\{q^g_i\,:\,i=0,\,1,...,\,N'\}$ where $N'=N-(M+1)$. We express this as 
\begin{align}\label{xnc}
x_n = g_n^t\{q^g_i\}
\end{align}
where the time dependence of the functions $g_n^t$ is due to that of the $f_\mu^t$. There is generally no unique way however, to eliminate $N-M$ of the $x_n$. A different elimination results in a different collection of functions $\{{g'}^t_n\}$ and different coordinates $q^{g'}_i$ such that $x_n = {g'}^t_n\{q^{g'}_i\}$. The sets of coordinates and velocities labelled by $g$ and $g'$ are related by a generally time-dependent bijective (invertible) point transformation; 
\begin{equation} \label{point}
\begin{split}
&q^{g'}_i = q^{g'}_i(\{q_j^g\},t), \\
& {\dot q}^{g'}_i = \sum_{j=0}^{N'} {\partial q_i^{g'} \over \partial q^g_j} {\dot q}_j^g + \partial_t q_i^{g'}(\{q_j^g\},t).
\end{split}
\end{equation}

Writing the Lagrangian $L$ in terms of the coordinates $\{q_i^g\}$ defines an explicitly time-dependent Lagrangian $L_g^t$ as
\begin{align}\label{LgL}
L\{x_n,{\dot x}_n\} = L\{g_n^t\{q_i^g\},{\dot g}^t_n\{q_i^g,{\dot q}_i^g\}\} =: L_g^t\{q_i^g,{\dot q}_i^g\}
\end{align}
where
\begin{align}\label{gdot}
{\dot g}^t_n= \sum_{j=0}^{N'} {\partial g_n^t \over \partial q^g_j} {\dot q}_j^g + \partial_t g^t_n
\end{align}
is seen to be linearly dependent on the velocities ${\dot q}_j$.  %Because of the additional term $\partial_t g_n^t$ on the right-hand-side of Eq.~(\ref{gdot}), 
The Lagrangian $L_g^t$ generally differs from any Lagrangian $L_g(t)$ that is obtained by phenomenologically introducing time dependence within the parameters of a time-independent Lagrangian $L_g$. 

Any two Lagrangian functions $L_g^t$ and $L_{g'}^t$ are distinct and are related using Eqs.~(\ref{point}) and (\ref{LgL}) as $L_g^t\{q_i^g,{\dot q}_i^g\} = L_{g'}^t\{q_i^{g'}\{q_i^g,t\},{\dot q}_i^{g'}\{q_i^g,{\dot q}_i^g,t\}\}$. Extending this formalism to deal with non-holonomic constraints is a difficult problem. We can however include some such cases by allowing $L_g^t\{q_i^g,{\dot q}_i^g\}$ and $L_{g'}^t \{q_i^{g'},{\dot q}_i^{g'}\} $ to differ by a total time-derivative of a function of the coordinates as
\begin{align}\label{LgLgp}
L_{g'}^t\{q^{g'}_j,{\dot q}^{g'}_j\} = L_g^t\{q^g_j,{\dot q}^g_j\} +{d\over dt}F^t_{gg'}\{q^g_j\}.
\end{align}
In this case the two Lagrangians, although not equal, remain equivalent.

Our treatment covers a broad class of systems, with two representative subclasses discussed in Sec.~\ref{examples}. The systems excluded are those subject to non-holonomic constraints that cannot be reduced to holonomic systems via gauge transformations. A classic example is a non-slipping rolling wheel. A description of such systems requires the non-variational Lagrange–d’Alembert principle, and although their external control could conceivably be of interest, this lies beyond the present scope.%Although the class of systems covered by our treatment is broad, with two broad example subclasses provided in Sec.~\ref{examples}, it is not all-encompassing. The systems lying beyond it are those with non-holonomic constraints that additionally cannot be reduced to a collection of holonomic constraints combined with gauge transformations. A well-known example is a non-slipping rolling wheel. Here, a rather different approach in the form of the Lagrange–d’Alembert principle is needed, but unlike Hamilton's, this principle is not variational. While important instances of external control could conceivably arise within such systems, this topic lies beyond the scope of the present article.

\subsection{The canonical formalism and the irrotational gauge}

Our aim now is to compare the results of introducing an explicit time dependence prior to  the construction of the Lagrangian versus afterwards, for instance, at the Hamiltonian level.

\subsubsection{Hamiltonian}\label{timedep}

Construction of the Hamiltonian proceeds in the usual way. We assume that the imposition of constraints has necessarily resulted in a non-degenerate Lagrangian, such that there are no identically zero canonical momenta. The latter are defined by
\begin{align}\label{can2}
p^g_j = {\partial L_g^t \over \partial {\dot q}^g_j}
\end{align}
where $L_g^t$ is given in Eq.~(\ref{LgL}). For different $g$ the canonical momenta are related by
\begin{align}
p^{g'}_i = \sum_{j=0}^{N'} p_j^g {\partial q^g_j \over \partial q^{g'}_i} +{\partial F^t_{gg'} \over \partial q_i^{g'}},
\end{align}
from which it follows that $\partial p^{g'}_i /\partial p^g_j = \partial q^g_j /\partial q^{g'}_i$, and therefore that the transformation $g\leftrightarrow g'$ is canonical. The velocities ${\dot q}^g_i$ and canonical momenta are in one-to-one correspondence, such that the former can be written as time-dependent functions of the canonical variables; ${\dot q}^g_j \equiv {\dot q}^g_j\{q^g_i,p^g_i,t\}$. The Hamiltonian is then defined as
\begin{align}\label{ham2}
H_g^t\{q^g_i,p^g_i\} := \sum_{j=0}^{N'} p^g_j {\dot q}^g_j\{q^g_i,p^g_i,t\}- L^t_g\{q^g_j,{\dot q}^g_j\{q^g_i,p^g_i,t\}\}
\end{align}
and it is explicitly time-dependent. For different $g$ the Hamiltonians are related by
\begin{align}\label{ham2b}
&H_{g'}^t\{q^{g'}_i,p^{g'}_i\} \nonumber \\ &= H_g^t\{q^g_i,p^g_i\} + \sum_{i=0}^{N'}p^{g'}_i\partial_t q_i^{g'}(\{q_j^g\},t) - \partial_t F^t_{gg'}\{q_i^g\}.
\end{align}
We have therefore constructed a physically unique canonical theory, including a description of any external control  that can be described using holonomic constraints and gauge transformations. %If preferred, one can express the right-hand side of Eq.~(\ref{ham2b}) entirely in terms of $g$-coordinates by noting that
%\begin{align}
%\sum_{i=0}^{N'} p^{g'}_i\,\partial_t q^{g'}_i(\{q_j^g\},t)=-\sum_{i=0}^{N'}\left[p_i^g+{\partial F_{gg'} \over \partial q_j^g}\right]\,\partial_t q_i(\{q_j^{g'}\},t).
%\end{align}

Upon quantisation the canonical variables become canonical operators and we consider dynamics in the Schr\"odinger picture meaning operators change only due to explicit time dependence. The transformation $g\leftrightarrow g'$ is implemented by an explicitly time-dependent unitary operator $U_{gg'}(t)$ as
\begin{align}
q^{g'}_i(\{q_j^g\},t) &= U_{gg'}(t)^\dagger q^g_i U_{gg'}(t), \label{t1}\\ 
p^{g'}_i(\{q_j^g,p_j^g\},t) &= U_{gg'}(t)^\dagger p^g_i U_{gg'}(t).\label{t2}
\end{align}
Each of the canonical operator sets (labelled by $g$ and $g'$) is explicitly time-dependent when expressed in terms of the other set. It is sufficient for us to consider separately the limiting cases of {\em i}) a pure point transformation, $F_{gg'}\equiv 0$, and {\em ii}) a pure ``gauge" transformation $q_i^g = q_i^{g'}$. A general treatment of point transformations [case {\em i})] can be found in Ref.~\cite{dewitt_point_1952}. We will restrict our attention to pure translations such that  $\partial q^g_i /\partial q^{g'}_j = \delta_{ij}$. This suffices to provide a clear example in Sec.~\ref{examples} of the significance of different treatments of explicit time dependence. The unitary transformation required to translate each coordinate $q_i$ by an arbitrary time-dependent function $c_i(t)$ is simply
\begin{equation} \label{Uggpt}
\begin{split}
&U_{gg'}(t) = \exp\left(iS^t_{gg'}\right),\\
&S_{gg'}^t = \sum_ {i=0}^{N'}c_i(t)p_i^{g}.
\end{split}
\end{equation}
Equations~(\ref{t1}) and (\ref{t2}) read $q_i^{g'}(\{q_j^g\},t)=q_i^g - c_i(t)$ and $p_i^{g'}= p_i^g$ respectively, while Eq.~(\ref{ham2b}) reads
\begin{align}\label{ham22}
H_{g'}^t\{q^{g'}_i,p^{g'}_i\} &= H_g^t\{q^g_i,p^g_i\} - \partial_t S_{gg'}^t .
\end{align}
Letting $q_i^g\equiv q_i$ and $p_i\equiv p_i^g$, and noting that $ \partial_t S_{gg'}^t = -i{\dot U}_{gg'}(t)U_{gg'}(t)^\dagger= i {\dot U}_{gg'}(t)^\dagger U_{gg'}(t)$ and $\partial_t [U_{gg'}(t) U_{gg'}(t)^\dagger] \equiv 0$, we see using Eqs.~(\ref{t1}) and (\ref{t2}) that the two Hamiltonians $H_g^t$ and $H_{g'}^t$ are equivalent;
\begin{align}\label{ham3}
H_{g'}^t\{q_i,p_i\} = U_{gg'}(t) H_{g}^t\{q_i,p_i\} U_{gg'}(t)^\dagger+i{\dot U}_{gg'}(t)U_{gg'}(t)^\dagger.
\end{align}

In the case {\em ii}) of a pure ``gauge" transformation, $q_i^{g'}=q_i^{g}$, the required unitary transformation is
\begin{equation} \label{Uggpt2}
U_{gg'}(t) = \exp\left(iF^t_{gg'}\right)
\end{equation}
such that Eqs.~(\ref{t1}) and (\ref{t2}) read $q_i^{g'}=q_i^g$ and $p^{g'}_i = p_i^g + {\partial F_{gg'}/\partial q_i^g}$ respectively, while Eq.~(\ref{ham2b}) reads
\begin{align}\label{ham23}
H_{g'}^t\{q^{g'}_i,p^{g'}_i\} &= H_g^t\{q^g_i,p^g_i\} - \partial_t F_{gg'}^t.
\end{align}
Letting $q_i^g\equiv q_i$ and $p_i\equiv p_i^g$, and applying the same manipulations that lead from Eq.~(\ref{ham22}) to Eq.~(\ref{ham3}), one again obtains Eq.~(\ref{ham3}). 

\subsubsection{Naive Hamiltonians and the irrotational gauge}\label{timeindep}

Let us now suppose that the theory does not possess any explicit time dependence through either the constraints $f_\mu =0$ or the gauge function $F_{gg'}$. The point transformation $q_i^{g'}=q_i^{g'}\{q_j^g\}$ is then also time-independent as are Eqs.~(\ref{t1}) and (\ref{t2}) for the quantum canonical transformation $g\leftrightarrow g'$. Equation~(\ref{ham2b}) becomes simply $H_{g'}\{q^{g'}_i,p^{g'}_i\} = H_g\{q^g_i,p^g_i\}$. Letting $q_i^g\equiv q_i$ and $p_i\equiv p_i^g$ one therefore obtains
\begin{align}\label{uhu}
H_{g'}\{q_i,p_i\} = U_{gg'}H_g\{q_i,p_i\} U_{gg'}^\dagger.
\end{align}
Suppose now that we introduce time dependence into the Hamiltonian $H_g$ in the Schr\"odinger picture as $H_g\to H_g(t)$. This could for example be the result of introducing $t$-dependence into the functions $g_n$ upon which $H_g$ depends, resulting in concurrent $t$-dependence of the $g_n'$ and $U_{gg'}$. However, if the functional forms of $H_g$, $H_{g'}$, and $U_{gg'}$ in terms of the canonical operators remain unchanged, then Eq.~(\ref{uhu}) will continue to hold in the form $H_{g'}(t)=U_{gg'}(t)H_g(t)U_{gg'}(t)^\dagger$. The Hamiltonians $H_{g}(t)$ and $H_{g'}(t)$ obtained in this way, are therefore not equivalent. An equivalence class of Hamiltonians can of course be constructed for each fixed $g$ by allowing $g'$ to vary as
\begin{align}\label{equiv}
{\mathscr S}_g = \{H_g^{g'}(t) = H_{g'}(t)+i{\dot U}_{gg'}(t)U_{gg'}(t)^\dagger\,:\, g'~{\rm variable}\},
\end{align}
but distinct such classes ${\mathscr S}_g$ and ${\mathscr S}_{g'}$ are not equivalent. Thus, introducing time dependence at the Hamiltonian level does not in general result in a unique (unambiguous) description.

The freedom to choose among the different descriptions labelled by $g$ will henceforth be called gauge-freedom. In general, one cannot guarantee the existence of a gauge for which
\begin{align}
X_g(t) := H_g^t - H_g(t) = 0\label{Xg}
\end{align}
where $H_g^t$ is obtained by assuming explicit time dependence from the outset as in the previous section. However, if such a gauge does exist, its naive model $H_g(t)$ is physically valid, because it coincides with $H_g^t$. We adopt the terminology proposed in Ref.~\cite{you_circuit_2019}  and call such gauges {\em irrotational}. Given any naive model $H_g(t)$, one can derive a correct model from it as $H_{g_{\rm irr}}(t)=U_{gg_{\rm irr}}(t)H_g(t)U_{gg_{\rm irr}}(t)^\dagger$, without any additional contributions arising from the time dependence of the rotation. Note however, that the naive $g$-gauge model $H_g(t)$ is not itself correct unless $g=g_{\rm irr}$. The correct $g$-gauge model, $H_g^t$, can be obtained from the irrotational gauge model by inserting $H_{g_{\rm irr}}^t\equiv H_{g_{\rm irr}}(t)$ into the right-hand-side of Eq.~(\ref{ham3}). Letting $U_{gg'}(t)=\exp\left[i{\cal G}_{gg'}^t\right]$, this gives an expression for $X_g(t)$ in terms of $g_{\rm irr}$ as
\begin{align}
X_g(t) := i{\dot U}_{g_{\rm irr}g}(t)U_{g_{\rm irr}g}(t)^\dagger = -\partial_t {\cal G}_{g_{\rm irr}g}^t.\label{Ugirr}
\end{align}
For the examples of a translation and a pure gauge transformation given in the previous section, we have ${\cal G}_{gg'}^t = S_{gg'}^t$ and ${\cal G}_{gg'}^t=F_{gg'}^t$ respectively.

\section{Application to light-matter systems}\label{examples}

We provide two example applications of the framework presented in Sec.~\ref{general}.

\subsection{Superconducting circuit with variable external flux}\label{squid}

 Ref.~\cite{you_circuit_2019} presents a systematic approach to quantising electrical circuits that contain loops threaded by {\em external} fluxes. The authors consider an arbitrary circuit consisting of $N$ elements spread across $M$ primitive loops $L_j$ (each not containing any further loops). Denoting the external flux threading loop $L_j$ by $\phi_j(t)$, and denoting the $n$'th branch flux by $x_n$, there are $M$ linear constraints expressible in the form $\Phi(t) = R{\bf x}$ where $\Phi(t) =(\phi_0(t),...,\phi_{M-1}(t))$, ${\bf x}=(x_0,...,x_{N-1})$, and $R$ is an $M\times N$ matrix. It is defined by $R_{jn}=1$ (respectively $R_{jn}=-1$), when $\phi_j(t)$ and $x_n \in L_j$ possess the same (respectively opposite) orientations, while $R_{jn}=0$ if $x_n\not\in L_j$. The $x_n$ are also related to the $N-M$ unconstrained but gauge-dependent fluxes ${\bf q}^g=(q_0^g,...,q^g_{N-M-1})$ using an $(N-M)\times N$ gauge matrix $g=G$, as ${\bf q}^g=G {\bf x}$. Defining ${\bf Q}^g := ({\bf q}^g,\Phi(t))$ one obtains ${\bf Q}^g=S{\bf x}$ where the $N\times N$ matrix $S = (G,R)$ is, by construction, invertible; ${\rm det}{S}\neq 0$, such that ${\bf x}=S^{-1}{\bf Q}^g$ \cite{you_circuit_2019}. This last expression is a particular instance of Eq.~(\ref{xnc}) in which the $x_n$ happen to be {\em linear combinations} of the $q^g_i$ and the time-dependent functions $\phi_j(t)$, with time-independent coefficients $S_{nm}$. Clearly then, the construction of a canonical description of a circuit threaded by external fluxes constitutes an example application of the general framework we have provided in Sec.~\ref{general}.

\begin{figure}[t]
\begin{minipage}{\columnwidth}
\begin{center}
\includegraphics[scale=0.5]{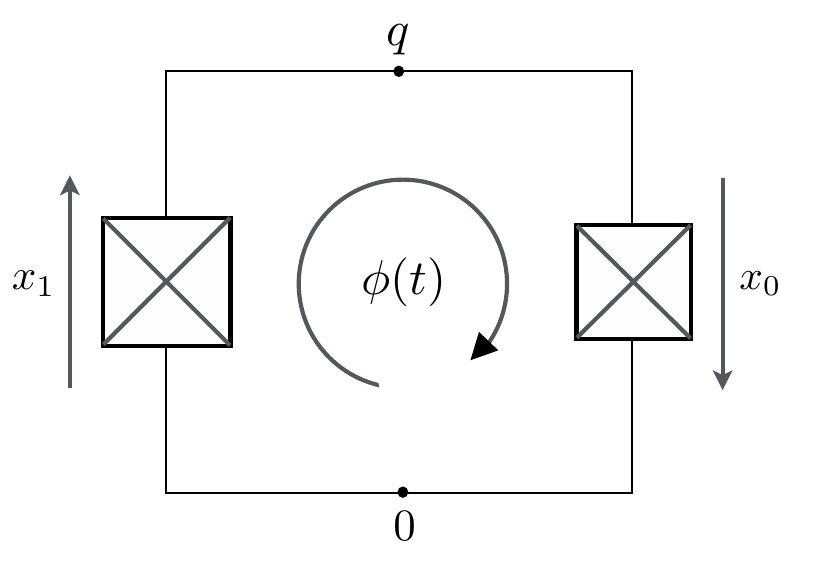}
\vspace*{-3mm}
\caption{Two Josephson junctions form a closed loop , $L_0$, which is threaded by an external flux $\phi(t)$. The right and left branch fluxes are $x_0$ and $x_1$, respectively.  Since $x_0$ and $x_1$ both possess the same orientation as $\phi(t)$, we have $R_{00} =1= R_{01}$. The lower node (0) is designated as ground, and the non-ground node flux is $q =G{\bf x}$. The latter linear combination of the $x_n$ is determined by the choice of spanning tree (gauge). }\label{circuit}
\end{center}
\end{minipage}
\end{figure}

By way of explicit demonstration, we choose the simplest example presented in Ref.~\cite{you_circuit_2019}; a superconducting circuit consisting of two junctions labelled right ($n=0$) and left ($n=1$) with corresponding capacitances $C_n$ and Josephson energies $E_{Jn}$. The junctions are connected via two nodes, one of which is the ground node. A generally time-dependent external flux $\phi(t)$ threads the circuit loop, as depicted in Fig.~\ref{circuit}.  Since there is only a single loop, both branch fluxes $x_n$ possess the same orientation as $\phi(t)$, such that
\begin{align}\label{cons}
\phi(t) =R{\bf x}=  x_0+x_1,\qquad R= (1~1)
\end{align}

In Ref.~\cite{you_circuit_2019}, the choice of spanning tree (gauge $g$) is encoded into two parameters, denoted therein by $m_1$ and $m_r$. These determine the circuit's non-ground flux coordinate $q$ in terms of the branch fluxes via the $1\times 2$ matrix $G=(m_r~ m_1)$, giving $q=m_r x_0+ m_1x_1$ \footnote{We are adopting the convention of using $q$ to denote the canonical {\em coordinate} (flux) as in Sec.~\ref{general}. This node flux should not be confused with the node charge, which is the momentum conjugate to $q$, and is denoted here by $p$.}. The standard gauge choices $(m_r~m_1)=(0~1)$ and $(m_r~m_1)=(-1~0)$, are particular cases that assign the flux $q$ entirely to the right and left branches respectively. One can alternatively express $q$ in terms of the sum $m_+=m_1+m_r$ and the difference $m_\Delta =m_1-m_r$. Here we will restrict our attention to the cases defined by $m_\Delta=1$ and $m_1 =\alpha \in [0,1]$. This suffices to reveal the important physical conclusions, while allowing us to encode the gauge choice $g$ more simply into the single parameter $g=\alpha$, the corresponding matrix being $G=(\alpha -1~\alpha)$. We continue to be able to smoothly interpolate between the standard gauges of choosing the right arm as spanning tree ($\alpha=0$) and choosing the left arm as spanning tree ($\alpha=1$).

In terms of $\alpha$ the non-ground node flux $q=m_1x_1+m_r x_0$ is 
\begin{align}\label{span}
q = G{\bf x}=\alpha x_1 - (1-\alpha)x_0,
\end{align}
so by letting ${\bf Q}=(q,\phi(t))$ one obtains ${\bf Q}=S{\bf x}$ where
\begin{align}
S=\begin{pmatrix}
\alpha-1 & \alpha \\
1 & 1
\end{pmatrix}.
\end{align}
One can eliminate $x_1$ and $x_0$ in favour of $q$ and $\phi(t)$ using Eqs.~(\ref{cons}) and (\ref{span}) to obtain [cf. Eq.~(\ref{xnc})]
\begin{align}
&x_0 = g_{\alpha,0}^t(q) = \alpha\phi(t) -q ,\\
&x_1 = g_{\alpha,1}^t(q) = (1-\alpha)\phi(t) +q,
\end{align}
 or more compactly, ${\bf x}=S^{-1}{\bf Q}$. Each different gauge $\alpha$ defines a different set of time-dependent functions $g_{\alpha,n}^t$.

The Lagrangian is the difference between the physical kinetic and potential energies;
\begin{align}
L\{x_n,{\dot x}_n\} = \sum_{n=0,1}\left[{1\over 2}C_n {\dot x}_n^2 - V_n(x_n) \right]
\end{align}
where
\begin{align}
V_n(x_n) = -E_{Jn} \cos \left({2\pi \over \varphi } x_n\right) 
\end{align}
is the potential energy of the $n$'th junction, in which $\varphi$ is the elementary flux quantum.
%%%%%%%%%%%%%%%%%%%%%%%%%%%%%%%%%%%%%%%%%%%%%%%%%%%%%%%%%%%%%%%%%%%%%%%%%%%%%%%%%%%%%%%%%%%%
%%
%%	F I G U R E S  S T A R T
%%
%%%%%%%%%%%%%%%%%%%%%%%%%%%%%%%%%%%%%%%%%%%%%%%%%%%%%%%%%%%%%%%%%%%%%%%%%%%%%%%%%%%%
%%%%%%%%%%%%%%%%%%%%%%%%%%%%%%%%%%%%%%%%%%
%%
%%	F I G U R E S  E N D
%%
%%%%%%%%%%%%%%%%%%%%%%%%%%%%%%%%%%%%%%%%%%%%%%%%%%%%%%%%%%%%%%%%%%%%%%%%%%%%%%%%%%%%
Let us first suppose that the external flux $\phi(t)\equiv \phi$ is steady, in which case the $g_{\alpha,n}^t \equiv g_{\alpha,n}$ are time-independent. Defining $L_\alpha(q,{\dot q}) : = L\{g_{\alpha,n}(q), {\dot g}_{\alpha,n}(q,{\dot q})\}$ and $p:=\partial L_\alpha /\partial {\dot q}$, one finds the corresponding Hamiltonian $H_\alpha$ to be
\begin{align}\label{Halph1}
H_\alpha(q,p)= {p^2\over 2C} +V_0(\alpha\phi-q) + V_1([1-\alpha]\phi+q)
\end{align}
where $C:=C_0+C_1$. The Hamiltonians of different gauges are unitarily related as in Eq.~(\ref{uhu}) by
\begin{align}\label{naive}
H_{\alpha'} = R_{\alpha\alpha'} H_{\alpha} R_{\alpha\alpha'}^\dagger
\end{align}
where
\begin{align}\label{R}
R_{\alpha\alpha'}:=e^{i S_{\alpha\alpha'}},\qquad S_{\alpha\alpha'}:=(\alpha-\alpha')p\phi.
\end{align}
The canonical theory so obtained is physically unique and unambiguous only for a steady flux. Indeed, if we denote by $H_\alpha(t)$ and $ R_{\alpha\alpha'}(t)$ the result of making the replacement $\phi \to \phi(t)$ within Eqs.~(\ref{Halph1}) and (\ref{R}) respectively, then Eq.~(\ref{naive}) becomes $H_{\alpha'}(t) = R_{\alpha\alpha'}(t) H_{\alpha}(t) R_{\alpha\alpha'}(t)^\dagger$, such that $H_\alpha (t)$ and $H_{\alpha'}(t)$ are not equivalent.

If one instead assumes that the external flux is time-dependent from the outset and defines $L_\alpha^t(q,{\dot q}) = L\{g^t_{\alpha,n}(q), {\dot g}^t_{\alpha,n}(q,{\dot q})\}$ and $p:=\partial L_\alpha^t /\partial {\dot q}$ as in Eqs.~(\ref{LgL}) and (\ref{can2}), 
one finds the Hamiltonian given by Eq.~(\ref{ham2}) to be
\begin{align}\label{til}
H_\alpha^t = H_\alpha(t) + X_\alpha(t)
\end{align}
where
\begin{align}\label{X}
X_\alpha(t):= \left(\alpha  -{C_1\over C}\right)p {\dot \phi}(t).
\end{align}
In writing Eq.~(\ref{til}) we have dropped the purely external kinetic term proportional to $\phi(t)^2$, which only affects the energy zero-point. In the Schr\"odinger picture we have immediately that
\begin{align}
X_{\alpha}(t)- X_{\alpha'}(t) =\partial_t S_{\alpha\alpha'}^t,
\end{align}
where
\begin{align}
\partial_t S_{\alpha\alpha'}^t:=(\alpha-\alpha')p{\dot \phi}(t) = -i{\dot R}_{\alpha\alpha'}(t)R_{\alpha\alpha'}(t)^\dagger,
\end{align}
and since $R_{\alpha\alpha'}(t)H_\alpha(t)R_{\alpha\alpha'}(t)^\dagger = H_{\alpha'}(t)$, it follows from Eq.~(\ref{til}) that
\begin{align}
H_{\alpha'}^t = R_{\alpha\alpha'}(t)H^t_\alpha R_{\alpha\alpha'}(t)^\dagger + i{\dot R}_{\alpha\alpha'}(t)R_{\alpha\alpha'}(t)^\dagger
\end{align}
as in Eq.~(\ref{ham3}). We have therefore now obtained a physically unique time-dependent description.

According to Eqs.~(\ref{Xg}) and (\ref{til}), the irrotational gauge for which the naive Hamiltonian $H_\alpha(t)$ is correct, is the solution $\alpha= \alpha_{\rm irr}$ of $X_\alpha(t)\equiv 0$, such that for any $\alpha$ we have $X_\alpha(t) = -\partial_t S_{\alpha_{\rm irr}\alpha}^t$, as in Eq.~(\ref{Ugirr}). According to Eq.~(\ref{X}) the required value is 
\begin{align}
\alpha_{\rm irr} := {C_1\over C_0+C_1}.
\end{align}
Notice that this is not a constant value independent of other model parameters, meaning that its value depends on the underlying physical context. The extremal right (left) gauge $\alpha=0$ ($\alpha=1$) becomes effectively irrotational only when the right (left) capacitance dominates the left (right) capacitance; $C_0 \gg C_1$ ($C_1 \gg C_0$). Concurrently, the node flux $q$ of the irrotational gauge,
\begin{align}
q = {1\over C}(C_1x_1 - C_0x_0),
\end{align}
is a quasi-relative flux; it is the difference between the physical fluxes after each one has been weighted by the ratio of the corresponding junction capacitance to the total capacitance.

\subsection{A moving atom}

As a second example we consider the description of time-dependent light-matter interactions \cite{stokes_ultrastrong_2021,settineri_gauge_2021,gustin_gauge-invariant_2023}. Introducing a phenomenological modulation $\mu(t)$ within the light-matter interaction Hamiltonians of different gauges results in non-equivalent models \cite{stokes_ultrastrong_2021,stokes_implications_2022}. This is another example of the general result given above in Sec.~\ref{timeindep}, and it necessitates a more fundamental description of the underlying control mechanism. In practice, shuttling material systems around is one of only a few methods available to modulate a light-matter interaction, and so we here use holonomic constraints to describe such motion explicitly. 

\subsubsection{System and Constraints}

For simplicity we consider a bound material system consisting of only two charges $e_n =\mp e,~n=0,\,1$, with masses $m_n$ and positions ${\bf r}_n$. The densities of charge and current are
\begin{align}
\rho({\bf x}) \equiv  \rho({\bf x},\{{\bf r}_n\}) = \sum_{n=0,1} e_n\delta({\bf x}-{\bf r}_n),\\
{\bf J}({\bf x}) \equiv {\bf J}({\bf x},\{{\bf r}_n\}) =\sum_{n=0,1} e_n{\dot {\bf r}}_n \delta({\bf x}-{\bf r}_n),
\end{align}
where here ${\bf x}$ denotes a point in space. These densities satisfy identically, the local conservation of charge
\begin{align}\label{cont}
\partial_\mu j^\mu = {\dot \rho} +\nabla \cdot {\bf J} = 0
\end{align}
where $j^0 =\rho$, $j^i=({\bf J})_i$, and repeated spacetime indices are summed. Equation~(\ref{cont}) does not have to be imposed as a constraint. To describe the gross motion of the two-charge system however, we introduce the holonomic constraint
\begin{align}\label{cofm}
m_0{\bf r}_0+m_1{\bf r}_1 = M{\bf R}(t)
\end{align}
where $M:=m_1+m_2$ and the centre-of-mass position ${\bf R}(t)$ is assumed to be explicitly time-dependent. The explicitly time-independent reduced coordinates for the material system are the components of the relative position ${\bf r}={\bf r}_0-{\bf r}_1$. The electromagnetic field is described by potential $A$ and the field tensor $F=dA$ possesses components $F_{\mu\nu} = \partial_\mu A_\nu - \partial_\nu A_\mu$. The longitudinal and transverse parts of the vector potential ${\bf A}$ are gauge-dependent and gauge-invariant respectively.

We encode gauge freedom into the real parameter $\alpha$ and choose the holonomic constraint \cite{stokes_identification_2021,woolley_charged_1999,woolley_reformulation_1974,woolley_non-relativistic_1975,stokes_implications_2022}
\begin{align}\label{cA}
\int d^3 x\, {\bm g}_\alpha({\bf x},{\bf x}',{\bf R}(t)) \cdot {\bf A}({\bf x}) = 0
\end{align}
in which ${\bm g}_\alpha = {\bm g}_{\rm L}+{\bm g}_{\alpha\rm T}$, where
\begin{align}
{\bm g}_{\rm L}({\bf x},{\bf x}',{\bf o}) := -\nabla {1\over 4\pi |{\bf x}-{\bf x}'|}
\end{align}
is independent of ${\bf o}$ and $\alpha$, and
\begin{equation}\label{gTt}
\begin{split}
 &{\bm g}_{\alpha\rm T}({\bf x},{\bf x}',{\bf o})  := -\alpha \int_{{\bf o}}^{\bf x'} d{\bf s}_\lambda\cdot  \delta^{\rm T}({\bf x}-{\bf s}),\\
&{\bf s}_\lambda :={\bf o} + \lambda[{\bf x}'-{\bf o}],\qquad \lambda \in [0,1].
\end{split}
\end{equation}
The transverse $\delta$-function possesses the Fourier representation
\begin{align}
\delta^{\rm T}_{ij}({\bf x}) = \int {d^3k \over (2\pi)^3}(\delta_{ij}-{\hat k}_i{\hat k}_j)e^{i{\bf k}\cdot {\bf x}},
\end{align}
which is seen to be the difference between $\delta_{ij}\delta({\bf x})$ and the corresponding longitudinal $\delta$-function $\delta_{ij}^{\rm L}({\bf x})$.
The constraint (\ref{cA}) is sufficiently general to yield the Coulomb ($\alpha=0$) and multipolar ($\alpha=1$) gauges as particular cases. The $\alpha$-gauge polarisation field
\begin{align}\label{Palph}
{\bf P}_\alpha({\bf x}) &:= -\int d^3 x\, {\bm g}_\alpha({\bf x},{\bf x}',{\bf R}(t))\rho({\bf x}')
\end{align}
satisfies $-\nabla \cdot {\bf P}_\alpha =-\nabla\cdot  {\bf P}_{\rm L} = \rho$ identically.  The $\alpha$-independent longitudinal part is the Coulomb gauge polarisation ${\bf P}_{\rm L} =\nabla \phi_{\rm Coul}$ where
\begin{align}\label{phic}
\phi_{\rm Coul}({\bf x}) : = -(\nabla^{-2}\rho)({\bf x}) = \int d^3 x' {\rho({\bf x}')\over 4\pi|{\bf x}-{\bf x}'|}
\end{align}
is the Coulomb potential. 

Line integral expressions of the type in Eq.~(\ref{gTt}) were introduced by Dirac relatively early on \cite{dirac_gauge-invariant_1955}. Most recently, Woolley has provided a discussion in Ref.~\cite{woolley_power-zienau-woolley_2020}. Equation~(\ref{gTt}) can be motivated by first noting that the Coulomb field $-{\bf E}_{\rm L}={\bf P}_{\rm L}$ of our neutral atom can be written
\begin{align}
{\bf P}_{\rm L}({\bf x})= q \int_{\bf o}^{\bf r} d{\bf s} \cdot \delta^{\rm L}({\bf x}-{\bf s})
\end{align}
independent of both the choice of origin ${\bf o}$ and specific path ${\bf s}$ that ends at ${\bf r}$. For the choices ${\bf o}={\bf R}(t)$ and ${\bf s}$ a straight-line, Eq.~(\ref{gTt}) gives for $\alpha=1$ the total multipolar polarisation
\begin{align}
{\bf P}_1({\bf x}) ={\bf d}\int_0^1 d\lambda \, \delta({\bf x}-{\bf R}(t)-\lambda[{\bf r}-{\bf R}(t)]).
\end{align}
Unlike the Coulomb field ${\bf P}_{\rm L}$, this field is strictly {\em localised inside the atom}, and since the canonical formalism yields the transverse photonic momentum ${\bf \Pi}=-{\bf E}-{\bf P}$ \cite{stokes_identification_2021,cohen-tannoudji_photons_1989}, we see that whether ${\bf \Pi}$ is locally connected to the local field ${\bf E}$ depends on the localisation properties of ${\bf P}$. In the Coulomb gauge ${\bf P}={\bf P}_{\rm L}=-{\bf E}_{\rm L}$ is non-local, and so therefore, is ${\bf \Pi}=-{\bf E}_{\rm T}=-{\bf E}+{\bf E}_{\rm L}$. By contrast, for any origin ${\bf o}$ and path ${\bf s}$ confined within the atom, the line-integral form of ${\bf g}_{\rm T}$ [Eq.~(\ref{gTt})] yields ${\bf \Pi}=-{\bf E}$ at all points outside the atom, providing the best possible transverse representation of the total electric field \cite{cohen-tannoudji_photons_1989}. The PZW straight line choice in Eq.~(\ref{gTt}) is particularly noteworthy because it enables a simple multipolar development of the polarisation and corresponding magnetisation fields. In particular, the interaction Hamiltonian can be expanded in a series of multipole moments, facilitating a quantum description that parallels classical multipolar radiation theory.

Healy showed that the line-integral ${\bf g}_{\rm T}$ does not in fact provide the most general partition of the charge and current densities into polarisation and magnetisation fields \cite{healy_representation_1977}. One could also consider imposing more general gauge constraints that, for example, are non-linear in ${\bf A}$, unlike Eq.~(\ref{cA}) \cite{chernyak_non-linear_2015}. Here, however, such generalisations offer little advantage: our simple $\alpha$-gauge formalism is the minimal extension needed to track how gauge choice affects an explicitly time-dependent canonical theory. %Notable non-trivial gauges ($\alpha\neq0,1$) include the Jaynes–Cummings gauge and irrotational gauges \cite{stokes_gauge_2019,stokes_ultrastrong_2021}.”

\subsubsection{Hamiltonian}\label{correctham}

We now proceed with the construction of the canonical description. The two time-dependent holonomic constraints, (\ref{cofm}) and (\ref{cA}), determine the constrained variables ${\bf r}_0, {\bf r}_1$, and ${\bf A}={\bf A}_{\rm T}+{\bf A}_{\rm L}$, in terms of two unconstrained variables, ${\bf r}={\bf r}_0-{\bf r}_1$ and ${\bf A}_{\rm T}$, and the time-dependent external vector ${\bf R}(t)$, as [cf. Eq.~(\ref{xnc})]
\begin{align}
{\bf r}_0 &= {\bf g}_0^t({\bf r}) :=  {\bf R}(t)+{m_1\over M}{\bf r},\label{clm1}\\
{\bf r}_1 &= {\bf g}_1^t({\bf r}) := {\bf R}(t)-{m_0\over M}{\bf r},\label{clm2}\\
%\end{align}
%The constraint~(\ref{cA}) can be used to eliminate the gauge-dependent longitudinal potential ${\bf A}_{\rm L}$ in favour of the gauge-invariant part ${\bf A}_{\rm T}$ and an $\alpha$-dependent function of ${\bf R}(t)$ as [cf. Eq.~(\ref{xnc})]
%\begin{equation}\label{clm3}
%\begin{split}
{\bf A}({\bf x}) &= {\bf g}_{2,\alpha}^t([{\bf A}_{\rm T}], {\bf x}) :=  {\bf A}_{\rm T}({\bf x}) + \nabla \chi_\alpha([{\bf A}_{\rm T}],{\bf x},{\bf R}(t)),\label{clm3} \\
\chi_\alpha &([{\bf A}_{\rm T}],{\bf x},{\bf R}(t)) := \int d^3 x'  {\bm g}_\alpha({\bf x}',{\bf x},{\bf R}(t)) \cdot {\bf A}_{\rm T}({\bf x}').
%\end{split}
\end{align}
Any function $F$ of the ${\bf r}_n$ and ${\bf A}_{\rm L}$ becomes explicitly time-dependent when written in terms of ${\bf r}$, ${\bf A}_{\rm T}$ and ${\bf R}(t)$, and we will henceforth abbreviate it as $F^t := F({\bf r}_1({\bf r},{\bf R}(t)),{\bf r}_2({\bf r},{\bf R}(t)),[{\bf A}_{\rm L}([{\bf A}_{\rm T}],{\bf R}(t))])$. Simple examples depending only on the ${\bf r}_n$ and their velocities include $\rho^t,~{\bf J}^t$ and ${\bf P}_\alpha^t$.

We must now construct a Lagrangian function of ${\bf r}$, ${\bf A}_{\rm T}$, and ${\bf R}(t)$, and their velocities. The standard QED Lagrangian assuming non-relativistic charges is
\begin{align}\label{Lem}
&L({\bf r}_1,{\bf r}_2, [A_\mu,(d A)_{\mu\nu}]) \nonumber \\ &={1\over 2}m_1{\dot {\bf r}}_1 + {1\over 2}m_2{\dot {\bf r}}_2 - \int d^3 x\left(j^\mu A_\mu + {1\over 4}F^{\mu\nu}F_{\mu\nu}\right) %\nonumber \\ =& {1\over 2}m_1{\dot {\bf r}}_1 + {1\over 2}m_2{\dot {\bf r}}_2 - V({\bf r}_1-{\bf r}_2)+ \int d^3 x \left[\rho\partial_t\chi^\alpha +{\bf J}\cdot {\bf A} +{1\over 2}({\bf E}_{\rm T}^2-{\bf B}^2)\right] \nonumber \\ =&
%{1\over 2}m{\dot {\bf r}} + {1\over 2}M{\dot {\bf R}} - V({\bf r})+ \int d^3 x \left[{\bf J}\cdot {\bf A}_{\rm T} -{d\over dt}{\bf P}^\alpha_{\rm T}\cdot {\bf A}_{\rm T} +{1\over 2}({\bf E}_{\rm T}^2-{\bf B}^2)\right] 
\end{align}
which is subject not only to the holonomic constraints (\ref{cofm}) and (\ref{cA}), but also to Gauss' law 
\begin{align}\label{gauss}
\nabla \cdot {\bf E} = \rho
\end{align}
in which the electric field ${\bf E} = -{\dot {\bf A}}-\nabla A_0$ depends on ${\dot {\bf A}}$. One way to arrive at the correct canonical quantum theory in this case, is to follow Dirac by systematically constructing Lie brackets consistent with the constraints~\cite{stokes_identification_2021,woolley_charged_1999,woolley_reformulation_1974,woolley_non-relativistic_1975,stokes_implications_2022}. However, the same final result can be obtained more straightforwardly by judiciously imposing the constraints at the outset. This yields an $\alpha$- and $t$-dependent Lagrangian function of the unconstrained coordinates and velocities, $L_\alpha^t[{\bf r},{\bf A}_{\rm T},{\dot {\bf r}},{\dot {\bf A}}_{\rm T}] $, which is of exactly the type appearing in Eq.~(\ref{LgL}). Subsequently, the procedure in Sec.~\ref{general} can be applied, including textbook quantisation \cite{stokes_identification_2021}. To this end note first that since ${\bf E}=-\partial_t {\bf A}_{\rm T}-\nabla \phi_{\rm Coul} = -\partial_t {\bf A}-\nabla A_0$ is manifestly gauge-invariant, Eq.~(\ref{clm3}) suffices to fix $A_0$ up to a constant as
\begin{align}\label{A0}
A_0 = \phi_{\rm Coul}-\partial_t \chi_\alpha.
\end{align}
Gauss' law (\ref{gauss}) implies further that ${\bf E}_{\rm L}= -{\bf P}_{\rm L}=-\nabla \phi_{\rm Coul}$, such that
\begin{align}
\int d^3 x \, {\bf E}_{\rm L}({\bf x})^2 = \int d^3 x\, \phi_{\rm Coul}({\bf x})\rho({\bf x}) =  -{q^2 \over 2\pi |{\bf r}|} =: 2V({\bf r})\label{elpc}
\end{align}
depends only on ${\bf r}$ and does not therefore depend explicitly on $t$. We have dropped here the infinite Coulomb self-energies of the individual charges.

Using Eqs.~(\ref{cont}), (\ref{A0}), and (\ref{elpc}), the Lagrangian in Eq.~(\ref{Lem}) can be written \cite{stokes_ultrastrong_2021}
\begin{align}\label{Lalph}
L_\alpha^t[{\bf r},{\bf A}_{\rm T},{\dot {\bf r}},{\dot {\bf A}}_{\rm T}]  :=& {1\over 2}m{\dot {\bf r}} + {1\over 2}M{\dot {\bf R}}(t) - V({\bf r}) \nonumber \\ &+ \int d^3 x \bigg[{1\over 2}\left({\dot {\bf A}}_{\rm T}^2-(\nabla \times {\bf A}_{\rm T})^2 \right) \nonumber \\ &+ {\bf J}^t\cdot {\bf A}_{\rm T} -{d\over dt}\left({\bf P}^t_{\alpha \rm T}\cdot {\bf A}_{\rm T}\right)\bigg],
\end{align}
where $m:=m_1m_2/M$. Lagrangians of different gauges are related as in Eq.~(\ref{LgLgp}) by
\begin{align}
L_{\alpha'}^t - L_\alpha^t = {dF^t_{\alpha\alpha'}\over dt}
\end{align}
where
\begin{align}\label{Salph}
F^t_{\alpha\alpha'} = \int d^3 x \left[{\bf P}^t_{\alpha\rm T}({\bf x})-{\bf P}^t_{\alpha'\rm T}({\bf x})\right] \cdot {\bf A}_{\rm T}({\bf x}).
\end{align}
Importantly, the interaction component of $L_\alpha^t$ given by the final line of Eq.~(\ref{Lalph}), includes ${\dot {\bf R}}(t)$-dependent terms coming from {\em both} ${\dot {\bf P}}_{\alpha\rm T}^t$ and ${\bf J}^t$, which are proportional to $\alpha$ and $\alpha$-independent respectively. It follows that the Coulomb gauge, $\alpha=0$, is not generally irrotational.

The correct Hamiltonian for the moving atom, $H_\alpha^t$, can now be found using $L_\alpha^t$ as in Eq.~(\ref{ham2}). We provide an explicit expression only for the simple example of an electric dipole, which most clearly serves to illustrate the significance of our treatment. A strong electric dipole approximation (EDA), also known as the long-wavelengths approximation, is defined as \cite{stokes_identification_2021,cresser_rate_2003}
\begin{align}
\rho^t({\bf x}) &= -{\bf d}\cdot \nabla\delta({\bf x}-{\bf R}(t)),\\
{\bf J}^t({\bf x}) &= {\dot {\bf d}}\delta({\bf x}-{\bf R}(t))-{\dot {\bf R}}(t)({\bf d}\cdot \nabla)\delta({\bf x}-{\bf R}(t)),\label{J2}\\
{\bf P}^t_{\alpha{\rm T}}({\bf x})&={\bf d}\cdot \delta^{\rm T}({\bf x}-{\bf R}(t))
\end{align}
where ${\bf d}=-e{\bf r}$. The ${\dot {\bf R}}(t)$-dependent but $\alpha$-independent term in Eq.~(\ref{J2}) for ${\bf J}^t$ is especially noteworthy. It is the EDA of the so-called {\em R\"ontgen} current associated with the gross translational motion of the atom \cite{craig_molecular_1998,baxter_canonical_1993,cresser_rate_2003,boussiakou_quantum_2002,lembessis_theory_1993,wilkens_spurious_1993,wilkens_significance_1994}. This accounts for the magnetization possessed by a moving electrically polarized system when observed in the lab frame, even though the same system appears unmagnetised to a comoving observer \cite{craig_molecular_1998}. Without the R\"ontgen current, the theory is fundamentally inconsistent, because local conservation of charge, Eq.~(\ref{cont}), fails to hold.

The construction of an unambiguous canonical theory now proceeds in an identical way to the example in Sec.~\ref{squid}. The canonical momenta conjugate to ${\bf r}$ and ${\bf A}_{\rm T}$ are denoted ${\bf p}$ and ${\bf \Pi}$ respectively. The Hamiltonian found from $L_\alpha^t$ via Eq.~(\ref{ham2}) can be written
\begin{align}\label{ham4}
&H^t_\alpha = H_\alpha(t)+X_\alpha(t).
\end{align}
Here $H_\alpha(t)$ is obtained from the replacement ${\bf R}\to {\bf R}(t)$ within the corresponding time-independent Hamiltonian $H_\alpha$ as
\begin{align}\label{Halphtn}
H_\alpha(t) =&  {1\over 2m}[{\bf p}+e(1-\alpha){\bf A}_{\rm T}({\bf R}(t))]^2 + V({\bf r}) \nonumber \\ &+  {1\over 2}\int d^3x \left[({\bf \Pi}+{\bf P}^t_{\alpha\rm T})^2+(\nabla \times {\bf A}_{\rm T})^2\right],
\end{align}
while $X_\alpha(t)$ is an additional term defined by
\begin{align}\label{X2}
X_\alpha(t) :=& -{\dot {\bf R}}(t)\cdot \left[({\bf d}\cdot \nabla_{\bf R}){\bf A}_{\rm T}({\bf R}(t))\right] \nonumber \\ &+\alpha ({\dot {\bf R}}(t)\cdot \nabla_{\bf R}){\bf d}\cdot {\bf A}_{\rm T}({\bf R}(t)).
\end{align}
In writing Eq.~(\ref{ham4}) we have again dropped the purely external kinetic energy. In the Schr\"odinger picture we have immediately that 
\begin{align}
X_{\alpha}(t)- X_{\alpha'}(t) =\partial_t F_{\alpha\alpha'}^t
\end{align}
where $F_{\alpha\alpha'}^t = (\alpha-\alpha') {\bf d}\cdot {\bf A}_{\rm T}({\bf R}(t))$ is the EDA of $F_{\alpha\alpha'}^t$ in Eq.~(\ref{Salph}), such that 
\begin{align}
\partial_t F_{\alpha\alpha'}^t &= (\alpha-\alpha')({\dot {\bf R}}(t)\cdot \nabla_{\bf R}) {\bf d}\cdot {\bf A}_{\rm T}({\bf R}(t)) \nonumber \\&=- i{\dot R}_{\alpha\alpha'}(t)R_{\alpha\alpha'}(t)^\dagger
\end{align}
in which $R_{\alpha\alpha'}(t) := \exp\left( iF_{\alpha\alpha'}^t\right)$. Since $H_{\alpha'}(t)=R_{\alpha \alpha'}(t)H_\alpha(t) R_{\alpha\alpha'}(t)^\dagger$, it follows from Eq.~(\ref{ham4}) that the Hamiltonians $H_\alpha^t$ and $H_{\alpha'}^t$ are related by
\begin{align}
H_{\alpha'}^t= R_{\alpha \alpha'}(t)H_\alpha^t R_{\alpha\alpha'}(t)^\dagger +i{\dot R}_{\alpha \alpha'}(t)R_{\alpha \alpha'}(t)^\dagger,
\end{align}
as in Eq.~(\ref{ham3}).

In the dipole gauge ($\alpha =1$) the term $X_\alpha(t)$ defines the well-known dipolar R\"ontgen interaction
\begin{align}\label{ront}
X_1(t) = -{\bf d}\cdot[{\dot {\bf R}}(t)\times {\bf B}({\bf R}(t))] 
\end{align}
where ${\bf B}=\nabla \times {\bf A}_{\rm T}$. In words, the dipole couples to an additional electric field generated by its gross motion within the lab-frame magnetic field.

The irrotational gauge is defined by the condition $X_\alpha(t)\equiv 0$ for which the existence of a solution, $\alpha_{\rm irr}$, clearly cannot be guaranteed. One also sees clearly why the Coulomb gauge ($\alpha=0$) is not generally irrotational; the first term in Eq.~(\ref{X2}) is non-vanishing for all $\alpha$. It results from the essential second term in Eq.~(\ref{J2}). which is the EDA of the R\"ontgen current. %The latter is a measurable physical property and is present in every gauge, whether or not the EDA, which preserves gauge invariance, is performed.
In certain situations, $\alpha_{\rm irr}$ does exist, and an example is given in Ref.~\cite{stokes_ultrastrong_2021} for the case of a dipole moving through a single Gaussian cavity mode. This is discussed further in what follows.

\section{Implications and discussion}\label{discuss}

\subsection{Modulated light-matter interactions}

%Our analysis results in conclusions that contradict those found elsewhere \cite{settineri_gauge_2021,gustin_gauge-invariant_2023,di_stefano_resolution_2019}. 
We now explain how our results relate to those within recent literature on the topic of external time-dependence in light-matter systems \cite{settineri_gauge_2021,gustin_gauge-invariant_2023,di_stefano_resolution_2019,stokes_ultrastrong_2021}. We begin by noting that Eq.~(\ref{clm3}) is independent of ${\bf r}$ and dependent on both $\alpha$ and ${\bf R}(t)$, but for $\alpha=0$ it is independent of ${\bf R}(t)$. 
%The authors of Ref.~\cite{gustin_gauge-invariant_2023} describe the Coulomb gauge ($\alpha=0$) as uniquely ``time-invariant". 
It has been argued that for describing time-dependent light-matter interactions, modulation $\mu(t)$ should be introduced in the Coulomb gauge via the transverse potential as ${\bf A}_{\rm T} \to {\bf A}_{\rm T}' =\mu(t){\bf A}_{\rm T}$, and only then can a gauge-fixing transformation be performed. Under this assumption the time-independent Hamiltonian $H_\alpha$ should be replaced by
\begin{align}\label{wrong}
H_0^\alpha(t) = H_\alpha(t) + {\tilde X}_\alpha(t)
\end{align}
where $H_\alpha(t)$ is obtained by modulating the light-matter coupling within $H_\alpha$ through $\mu(t)$, and
\begin{align}
{\tilde X}_\alpha(t) : ={\dot \mu}(t) \int d^3 x {\bf P}_{\alpha\rm T}({\bf x}) \cdot {\bf A}_{\rm T}({\bf x})\label{wrong2}
\end{align}
where ${\bf P}_{\alpha\rm T}$ is defined by the time-independent case (${\bf R}(t)\equiv {\bf R}$) of Eq.~(\ref{Palph}). The Hamiltonian in Eq.~(\ref{wrong}) can be written $H_0^\alpha(t)= R_{0\alpha}(t)H_0(t) R_{0\alpha}(t)^\dagger +i{\dot R}_{0\alpha}(t)R_{0\alpha}(t)^\dagger$ where
\begin{align}\label{R0}
&R_{\alpha \alpha'}(t) :=\nonumber \\ &\exp\left[i\mu(t) \int d^3 x \left[{\bf P}_{\alpha\rm T}({\bf x})-{\bf P}_{\alpha'\rm T}({\bf x})\right] \cdot {\bf A}_{\rm T}({\bf x})\right].
\end{align}
The set $\{H_0^\alpha(t)\}$ is seen to be the Coulomb gauge equivalence class ${\mathscr S}_0$ defined in the manner described in Sec.~\ref{timeindep} via Eq.~(\ref{equiv}). %Equation (\ref{wrong}) above is Eq.~(80c) in Ref.~\cite{gustin_gauge-invariant_2023} (see also Eq.~(77) in Ref.~\cite{gustin_gauge-invariant_2023}), and the $\alpha=1$ case of Eq.~(\ref{wrong}) above is Eq.~(31) in Ref.~\cite{settineri_gauge_2021}.
The authors of Refs.~\cite{settineri_gauge_2021,gustin_gauge-invariant_2023} conclude that introducing $\mu(t)$ does not result in a different physical theory for each different $\alpha$, because only the class ${\mathscr S}_0$ is physically valid. Upon examining the physical properties of the canonical subsystems in different gauges it becomes clear that this conclusion cannot be correct in general.

Since ${\tilde X}_0\equiv 0$ [cf.~Eq.~(\ref{Ugirr})], the prescription (\ref{wrong}) amounts to the {\em assumption} that the irrotational gauge is always the Coulomb gauge. Physically, Eq.~(\ref{wrong}) constitutes the assumption that the electric field ${\bf E}' = -{\dot {\bf A}}'_{\rm T}-\nabla \phi_{\rm Coul}$ becomes explicitly time-dependent only through the transverse part ${\bf E}'_{\rm T}=-{\dot {\bf A}}'_{\rm T}$, which equals the field momentum in the Coulomb gauge. Therein however, the intra-atomic Coulomb energy does not exhaust the contribution of the Coulomb field, ${\bf E}_{\rm L} =-\nabla \phi_{\rm Coul}$, which according to Eq.~(\ref{phic}) responds instantaneously at ${\bf x}$ to changes in $\rho$ at any other point ${\bf x}'$ \cite{power_coulomb_1959}. Both ${\bf A}_{\rm T}$ and $-{\dot {\bf A}}_{\rm T}={\bf E}_{\rm T} = {\bf E}-{\bf E}_{\rm L}$ possess instantaneous non-local static contributions external to the atom. In contrast, modulation in the multipolar gauge occurs entirely inside the atom through a {\em local} field ${\bf E}+{\bf P}_1$. Clearly then, the gauge within which modulation is implemented implicitly determines the assumptions that are being made about the spacetime {\em localisation} properties of the external control. A more detailed discussion of these points can be found in Ref.~\cite{stokes_implications_2022}.

%At the very least, they should not be {\em arbitrarily favoured} over any other subsystems, as is implied by the generally incorrect prescription (\ref{wrong}). %Some further remarks are offered in the appendix.

\subsection{Beyond phenomenological modulation}

When considering more fundamental descriptions of the external control the prescription (\ref{wrong}) is immediately seen to fail because the gauge-fixing constraint will not generally be the only explicitly time-dependent one. For example, a time-dependent external flux allows modulation of resonator-qubit coupling \cite{huang_universal_2018,lu_universal_2017}, but external flux generally defines a separate constraint, such as Eq.~(\ref{cons}), from which it follows that the irrotational gauge is not fixed.

The shuttling of material systems is also one of the few methods available to modulate a light-matter interaction, but in this case the correct theory will very obviously include at least one other time-dependent constraint, namely, Eq.~(\ref{cofm}). Indeed, without Eq.~(\ref{cofm}) no time dependence can occur at all, because Eq.~(\ref{cofm}) defines ${\bf R}(t)$, which then results in time dependence of the gauge-fixing constraint. %Any term ${\tilde X}_\alpha(t)$ that is proportional to $\alpha$ will fail to coincide with the correct $X_\alpha(t)$. 
In particular, Eq.~(\ref{clm3}) is necessarily accompanied by the ${\bf R}(t)$-dependent equations~(\ref{clm1}) and (\ref{clm2}) yielding an $\alpha$-independent term that when combined with any $\alpha$-dependent terms gives $X_\alpha(t)$.
%A physically consistent description of dipolar motion in and out of a single Gaussian cavity mode was obtained in Ref.~\cite{stokes_ultrastrong_2021} following the construction of the correct Hamiltonian $H_\alpha^t$ in Eq.~(\ref{ham4}). It was demonstrated that in certain scena rios this description can indeed be expressed entirely in terms of a time-dependent coupling function $\mu(t)$. However, 
This term reduces to ${\tilde X}_\alpha(t)$ only if the contribution arising from the ${\dot {\bf R}}(t)$-dependent R\"ontgen current in Eq.~(\ref{J2}) is removed from it by hand.

In short, the assumption that the Coulomb gauge is necessarily irrotational is equivalent to the incorrect physical assumption that there is necessarily no R\"ontgen current. This yields ${\tilde X}_\alpha(t)$ in place of the correct result $X_\alpha(t)$. The R\"ontgen-current contribution, $X_\alpha(t)-{\tilde X}_\alpha(t)$, that is missing in $H_0^\alpha(t)\in {\mathscr S}_0$, is manifestly gauge-invariant and measurable. Within a correct theory, it is present in every gauge, including the Coulomb gauge, and this does not depend on any approximations such as the EDA. A theory without the R\"ontgen current is fundamentally inconsistent [it violates Eq.~(\ref{cont})] and it will fail to predict the correct physical behaviour of even the most elementary processes \cite{baxter_canonical_1993,cresser_rate_2003,boussiakou_quantum_2002,lembessis_theory_1993,wilkens_spurious_1993,wilkens_significance_1994}.

The trivial possibility of constructing an equivalence class ${\mathscr S}_\alpha$ from each $H_\alpha(t)$ was noted in Ref.~\cite{stokes_ultrastrong_2021}.  The {\em ab initio} selection of one fixed class, such as ${\mathscr S}_0$, as being preferred, would be arbitrary and as such unjustified. This is why the construction of a first-principles description was deemed necessary in Ref.~\cite{stokes_ultrastrong_2021}. Equating its result, $H_\alpha^t$, with the naive model $H_\alpha(t)$ removes the arbitrariness by determining the correct model $H_{\alpha_{\rm irr}}(t)$ whenever $\alpha_{\rm irr}$ exists, or else revealing that none of the classes ${\mathscr S}_\alpha$ are strictly valid. It was shown in Ref.~\cite{stokes_ultrastrong_2021} using the correct $X_\alpha(t)$ in Eq.~(\ref{X2}), that in some cases $\alpha_{\rm irr}$ does indeed exist. Specifically, when considering a harmonic dipole aligned with a Gaussian cavity mode's polarisation, which makes an angle $\theta$ with the constant motion ${\dot {\bf R}}(t)$ through the cavity (orthogonal to the cavity mode wavevector), one obtains $X_\alpha(t) = -e{\dot \mu}(t){\bf r}\cdot {\bf A}_{\rm T}({\bf 0})[\alpha-\cos^2 \theta]$ where $\mu(t) = \varphi({\bf R}(t))$ and $\varphi$ is the Gaussian mode function. The prescription (\ref{wrong}) meanwhile yields the incorrect result ${\tilde X}_\alpha(t) = -e\alpha{\dot \mu}(t){\bf r}\cdot {\bf A}_{\rm T}({\bf 0})$. As expected, the irrotational gauge $\alpha_{\rm irr} = \cos^2\theta$ is not a constant and vanishes only when $\theta=\pi/2$. In other words, which of the classes ${\mathscr S}_\alpha$ is the correct one, ${\mathscr S}_{\alpha_{\rm irr}}$, depends on the microscopic context.

\subsection{Natural lineshape}

Another well-known problem that is treatable using a phenominological modulation $\mu(t)$ but for which the prescription (\ref{wrong}) fails, is that of the spectrum of spontaneous emission (natural lineshape). An arbitrary-gauge but otherwise standard description is given in Appendix A. A bare excited dipole in the vacuum, $\ket{0,e}$, transitions into lower states with the emission of a photon. The lineshape is defined as the total spectral density of photons in the long-time limit. Since the vector $\ket{0,e}$ and the photon number operator represent a different physical state ${\cal I}_\alpha$ and observable ${\cal O}_\alpha$ in each different gauge $\alpha$, each spectrum $S_\alpha(\omega) \sim \langle {\cal O}_\alpha \rangle_{{\cal I}_{\alpha}(t=\infty)}$ refers to a different physical (gauge-invariant) property, and each would have to be measured using a different experiment \cite{stokes_implications_2022}.% The Schr\"odinger picture observable ${\cal O}_\alpha$ is represented in the gauge $\alpha'$ by the operator $R_{\alpha\alpha'}n_\lambda({\bf k})R_{\alpha\alpha'}^\dagger$ and the state ${\cal I}_\alpha(t)$ at time $t$ is represented in the gauge $\alpha'$ by the vector $ R_{\alpha\alpha'}U^\alpha(t)\ket{e,0}$. The same unique physical prediction $\langle{\cal O}_\alpha \rangle_{{\cal I}_\alpha(t)}$ is obtained in every gauge $\alpha'$, and this is true for every one of the distinct physical predictions labelled by $\alpha$.

The irrotational gauge concept provides additional insight into the problem, which as was noted by Milonni {\em et al.} \cite{milonni_natural_1989}, can be cast in terms of a modulated interaction. Specifically, the initial vector $\ket{0,e}$ uniquely represents an energy eigenstate provided the light-matter interaction is only switched-on at the initial time. The simplest assumption of a sudden switch-on $\mu(t) = \theta(t)$ (Heaviside step) within the gauge $\alpha$, yields a model $H_\alpha(t)$ that coincides with the time-independent Hamiltonian $H_\alpha$ for $t>0$, and therefore yields the same spectrum $S_\alpha(\omega)$ as the time-independent theory. The gauge-relativity of $S_\alpha(\omega)$ can now instead be understood by noting that introducing $\mu(t)$ constitutes a different physical assumption in each different gauge $\alpha$, such that the resulting naive models $H_\alpha(t)$ are not equivalent and describe different experiments \cite{stokes_ultrastrong_2021,stokes_implications_2022}.

~Lamb {\em et al.} \cite{lamb_matter-field_1987} (see also Refs.~\cite{rzazewski_equivalence_2004,funai_p_2019,scully_quantum_1997}) argue that the dipole gauge provides the relevant subsystem definitions, particularly with regard to early spectroscopic experiments \cite{lamb_fine_1950,lamb_fine_1951,lamb_fine_1952}. %Accordingly, correct predictions are to be obtained in any other gauge $\alpha \neq 1$ by using the appropriately transformed operators and vectors \cite{lamb_matter-field_1987,rzazewski_equivalence_2004,funai_p_2019}.
Essentially the same argument was originally made by Power and Zienau, who noted that the Coulomb gauge photonic operator $a^0_\lambda(t,{\bf k})$ suffers from static contributions characterised by a pole at $\omega=0$. This infrared divergent behaviour is obviously important in the far-field limit \cite{woolley_foundations_2022}, and is needed to exactly cancel the non-local Fourier components of instantaneous inter-atomic Coulomb interactions that are explicit within the Coulomb gauge \cite{power_coulomb_1959,craig_molecular_1998,cohen-tannoudji_photons_1989}. The PZW transformation $R_{01}$ yields a photonic operator $a^1_\lambda(t,{\bf k})$ (dipole gauge) that by design does not possess static admixtures \cite{power_coulomb_1959}.  As a result, in the dipole (and more generally multipolar) gauge all interactions occur through a local and properly retarded field ${\bf \Pi}=-{\bf E}-{\bf P}_1$ and there are no direct interatomic Coulomb interactions \cite{craig_molecular_1998,cohen-tannoudji_photons_1989}.

If, when introducing a modulation $\mu(t)$, the naive $\alpha$-gauge Hamiltonian $H_\alpha(t)$ is found to be correct, which is to say, the fixed gauge $\alpha$ is found to be irrotational, then the correct Hamiltonian of any other gauge $\alpha'$ is $H_\alpha^{\alpha'}(t)= R_{\alpha\alpha'}(t)H_\alpha(t) R_{\alpha\alpha'}(t)^\dagger +i{\dot R}_{\alpha\alpha'}(t)R_{\alpha\alpha'}(t)^\dagger$. The equivalence class ${\mathscr S}_\alpha = \{H_\alpha^{\alpha'}(t) \}$ generated in this way is as in Sec.~\ref{timeindep}. The Hamiltonian $H_\alpha^{\alpha'}(t)$ can be written
\begin{align}\label{right}
H_\alpha^{\alpha'}(t) = H_{\alpha'}(t) + X_{\alpha'}(t,\alpha)
\end{align}
where $H_{\alpha'}(t)$ is obtained by modulating the light-matter coupling within $H_{\alpha'}$ through $\mu(t)$, and
\begin{align}
X_{\alpha'}(t,\alpha) &: = i{\dot R}_{\alpha\alpha'}(t)R_{\alpha\alpha'}(t)^\dagger \nonumber \\ &= {\dot \mu}(t) \int d^3 x \left[{\bf P}_{\alpha'\rm T}({\bf x})-{\bf P}_{\alpha\rm T}({\bf x})\right] \cdot {\bf A}_{\rm T}({\bf x}) \nonumber \\ &= {\tilde X}_{\alpha'}(t)-{\tilde X}_\alpha(t)\label{right2}
\end{align}
such that $X_{\alpha}(t,0)={\tilde X}_\alpha(t)$. The prescription defined by Eqs.~(\ref{wrong}) and (\ref{wrong2}) is obtained by letting $\alpha=0$ in Eqs.~(\ref{right}) and (\ref{right2}), which constitutes the assumption that the irrotational gauge is always the Coulomb gauge.

If by comparison with experiments so far one takes the spectrum $S_1(\omega)$ as correct, then one concludes that the dipole gauge is irrotational. The assumption that $\mu(t)$ should be introduced in the Coulomb gauge via ${\bf A}_{\rm T}\to\theta(t){\bf A}_{\rm T}$ as in Eq.~(\ref{wrong}), that is, that the Coulomb gauge is irrotational, is then invalidated \cite{milonni_natural_1989}. Use of any member of the Coulomb gauge equivalence class ${\mathscr S}_0$ defined by Eq.~(\ref{wrong}) results in the spectrum $S_0(\omega)$. More generally, using any member of the equivalence class  ${\mathscr S}_\alpha$ defined by Eq.~(\ref{right}) results in the spectrum $S_\alpha(\omega)$. Details are given in Appendix A. 

It is important to note that, conceivably, different experiments could be constructed that grant access to different spectra $S_\alpha(\omega),~ \alpha\neq 1$. This aspect of the topic is somewhat nuanced and more detailed discussions can be found in, for example, Refs. \cite{stokes_implications_2022,stokes_gauge_2013,davidovich_theory_1980}. What is clear however, is that specifying a single gauge as the universally correct one within which to introduce a time-dependent modulation $\mu(t)$, is unsustainable. For describing an experiment that measures the spectrum $S_\alpha(\omega)$, the gauge $\alpha$ is irrotational and the correct equivalence class is ${\mathscr S}_\alpha$. %The same spectrum is obtained in any other gauge $\alpha'$ by generating the correct $\alpha'$ gauge model from $H_\alpha(t)$ using Eq.~(\ref{ham3}). This constitutes making the same physical assumption about modulation in every other gauge $\alpha'$.
The lineshape example demonstrates that the Coulomb gauge light and matter subsystems possess characteristics that would seem to render them particularly poorly suited for representing the physical subsystem preparation, control, and measurement protocols that are realised in actual experiments. When described properly then, simple physical examples show that time-dependent light-matter interactions are indeed gauge-relative \cite{stokes_ultrastrong_2021}. %It behoves us to address directly the various claims to the contrary made in Refs.~\cite{settineri_gauge_2021,gustin_gauge-invariant_2023,di_stefano_resolution_2019}, for which we refer the reader to Appendix B. 

\section{Conclusions}\label{conc}

We have presented a general framework for the construction of time-dependent Hamiltonian and Lagrangian theories with holonomic constraints. A main aim in doing so has been to fully understand the recently debated subject of introducing time dependence in quantum light-matter systems \cite{stokes_ultrastrong_2021,stokes_implications_2022,settineri_gauge_2021,gustin_gauge-invariant_2023,di_stefano_resolution_2019}. We have shown that in general a unique canonical theory is obtained only if time dependence is present within the constraints from the outset.

We have provided a general definition of the irrotational gauge as that within which a correct theory is also obtained when introducing time dependence later at the Hamiltonian level. The existence of such a gauge cannot be guaranteed, but several important examples certainly can be found. The idea is useful because it provides a generally simpler theory that is nevertheless correct. We have presented diverse examples within a unified framework \cite{you_circuit_2019,stokes_ultrastrong_2021}, showing in particular, that the Coulomb and irrotational gauges do not in general coincide. We have explained precisely how our results relate to those in existing literature on external time-dependence of light-matter systems %and why this contradicts treatments and conclusions in Refs.~
\cite{settineri_gauge_2021,gustin_gauge-invariant_2023,di_stefano_resolution_2019,stokes_ultrastrong_2021}.%, while reaffirming the prior treatment and conclusions in Ref.~\cite{stokes_ultrastrong_2021}.

  Since external control is necessary for next generation quantum technologies, explicit time-dependence is an increasingly prevalent feature of theoretical descriptions. In addition to the atom-cavity and superconducting circuit systems we have considered here, our general framework may apply to various other contexts. For example, in guiding further developments in the quantum theory of time-varying media, for which a simple, consistent Lagrangian formulation has recently been presented \cite{horsley_macroscopic_2025}, but which is yet to be extended to include a description of guest quantum emitters.

\section*{APPENDIX}

\section*{Appendix A}

We here provide a more detailed description of the natural lineshape problem. When adopting a passive view of gauge-fixing transformations the photon annihilation operators of different gauges are related using
\begin{align}\label{phrel1}
a^\alpha_{\lambda}(t,{\bf k}) = R^\dagger_{0\alpha} a^0_{\lambda}(t,{\bf k})R_{0\alpha} = a^0_{\lambda}(t,{\bf k}) - {i\alpha {\bf e}_\lambda ({\bf k}) \cdot{\bf d}(t) \over \sqrt{2\omega(2\pi)^3}}
\end{align}
where ${\bf e}_\lambda ({\bf k})$ is a unit polarisation vector orthogonal to ${\bf k}$. The lineshape can be defined as the total spectral density of photons, $n^\alpha_\lambda(t,{\bf k}) := a^\alpha_{\lambda}(t,{\bf k})^\dagger a^\alpha_{\lambda}(t,{\bf k})$, in the long-time limit;
\begin{align}
S_\alpha(\omega) %&= \lim_{t\to\infty} \int d\Omega \sum_\lambda |b_\lambda^\alpha(t,{\bf k})|^2 \nonumber \\ & = 
 &= \lim_{t\to\infty} \int d^3 k'  \sum_\lambda \langle n^\alpha_\lambda(t,{\bf k}')\rangle_{e,0}\delta(\omega-\omega')\label{ls} \\
 &= {\Gamma\over 2\pi} {(\omega /\omega_{eg}^3)[(1-\alpha)\omega_{eg}+\alpha\omega]^2 \over (\omega-\omega_{eg})^2+\Gamma^2/4}\label{ls2} \\
 &= \left[\alpha+(1-\alpha){\omega_{eg}\over \omega}\right]^2 S_1(\omega)
\end{align}
where the integration is over all photon momenta and the summation is over polarisations. For simplicity we have assumed that $e$ is the first excited level of the dipole and we have denoted the transition frequency to the bare ground level by $\omega_{eg}$. The result coincides with that obtained using the formal theory of radiation damping \cite{stokes_implications_2022} and it generalises to an arbitrary gauge $\alpha$ the well-known dipole and Coulomb gauge spectra found by various authors \cite{lamb_matter-field_1987,milonni_natural_1989,power_time_1999}. Equation (\ref{ls2}) is obtained by making the Markov and rotating-wave (Weisskopf-Wigner) approximations, within which the dipole moment decays exponentially and only the number conserving emission of a single real photon is permitted.

In any gauge the photon operator can be partitioned into vacuum and source parts as $a^\alpha_\lambda(t,{\bf k}) = a_{\lambda,\rm vac}^\alpha(t,{\bf k})+ a^
\alpha_{\lambda,s}(t,{\bf k})$ where $a_{\lambda,\rm vac}^\alpha(t,{\bf k}) = a^\alpha_\lambda(0,{\bf k})e^{-i\omega t}$, and the vacuum component does not contribute to the $\alpha$-gauge vacuum average of the normally ordered product appearing in Eq.~(\ref{ls}). Crucially however, vacuum-source partitions are gauge-relative \cite{stokes_implications_2022}. Eq~(\ref{phrel1}) implies that
\begin{align}\label{phrelvac}
a^\alpha_{\lambda,\rm vac}(t,{\bf k}) - a^1_{\lambda,\rm vac}(t,{\bf k}) &= i(1-\alpha) {{\bf e}_\lambda ({\bf k}) \cdot{\bf d}(0) \over \sqrt{2\omega(2\pi)^3}}e^{-i\omega t} \nonumber \\ &=:(1-\alpha)\delta_{\lambda}(t,{\bf k}).
\end{align}
and in turn that
\begin{align}
&a^\alpha_{\lambda,s}(t,{\bf k}) -a^1_{\lambda,s}(t,{\bf k})=
(1-\alpha) \left[{i{\bf e}_{\lambda}({\bf k}) \cdot {\bf d}(t) \over \sqrt{2\omega(2\pi)^3}} - \delta_{\lambda}(t,{\bf k})\right].\label{stat}
\end{align}
Since in the Weisskopf-Wigner approximation the dipole moment decays exponentially, the ${\bf d}(t)$-dependent term in Eq.~(\ref{stat}) vanishes for $t\to \infty$. The vacuum difference $\delta_\lambda(\infty,{\bf k})$ is non-vanishing however, so the lineshape $S_1(\omega)$ is only obtained for $\alpha\neq 1$ if $\delta_\lambda(t,{\bf k})$ is set to zero by hand \cite{milonni_natural_1989,stokes_implications_2022}. %This finding has been rediscovered very recently using a model obtained by rotating a truncated dipole gauge (Rabi) model \cite{gustin_gauge_2025}.
It is important to note that $\delta_\lambda(t,{\bf k})$ is time-dependent and non-vanishing for all times, so it is not merely a difference in {\em initial} conditions for the $a_\lambda^\alpha(t,{\bf k})$. Moreover, since even very early experiments are able to resolve the difference between $S_0(\omega)$ and $S_1(\omega)$, removing $\delta_\lambda^0(t,{\bf k})$ from $a^0_{\lambda,s}(t,{\bf k})$ cannot be justified as an approximation. Indeed, in Eq.~(\ref{phrelvac}) the initial dipole moment operator ${\bf d}(0)$, which is nothing but the Schr\"odinger picture operator, is necessarily non-vanishing. This is true classically as well; the charges comprising the dipole cannot be taken as identically coincident, despite their separation typically being small compared with resonant transition wavelengths (EDA). Imposing $\delta_\lambda^\alpha(t,{\bf k}) =0$ within Eq.~(\ref{stat}) is simply an ad hoc means by which to force the equality $a^\alpha_{\lambda,s}(\infty,{\bf k})=a^1_{\lambda,s}(\infty,{\bf k})$, thereby implementing within the gauge $\alpha$ the assumption that the dipole gauge subsystems are those most physically relevant \cite{power_coulomb_1959,lamb_matter-field_1987,rzazewski_equivalence_2004,funai_p_2019,scully_quantum_1997}. 

The equivalent treatment of the problem that instead assumes a sudden interaction switch-on, $\mu(t) = \theta(t)$ (Heaviside step), reveals that using any member of the Coulomb gauge equivalence class ${\mathscr S}_0$ defined by Eq.~(\ref{wrong}) results in the incorrect spectrum $S_0(\omega)$. To see this first note that for $\mu(t)=\theta(t)$ the additional term ${\tilde X}_\alpha(t)$ in Eq.~(\ref{wrong2}) is for a dipole at ${\bf 0}$ given by ${\tilde X}_\alpha(t) = \alpha\delta(t){\bf d}\cdot {\bf A}_{\rm T}({\bf 0})$, where ${\dot \mu}(t) =\delta(t)$. For a smoother switching function ${\dot \mu}(t)$ would be concurrently more smoothly peaked around $t=0$. The equations of motion found using the arbitrary member $H_0^\alpha(t)$ of ${\mathscr S}_0$ can be integrated from any initial time $t_0 <0$ at which $\mu$ vanishes, and for the case $\mu(t)=\theta(t)$ one obtains the photonic source operator
\begin{align}\label{wrong3}
a^\alpha_{0,\lambda,s}(t,{\bf k}) = a^\alpha_{\lambda,s}(t,{\bf k}) - \alpha\delta_\lambda(t,{\bf k})
\end{align}
where the second term is due to ${\tilde X}_\alpha(t)$. It is easily shown \cite{stokes_gauge-relativity_2023} that $a^\alpha_{\lambda,s}(t,{\bf k}) =(1-\alpha)a^0_{\lambda,s}(t,{\bf k}) + \alpha a^1_{\lambda,s}(t,{\bf k})$ from which it follows using Eq.~(\ref{stat}) that
\begin{align}
a^\alpha_{0,\lambda,s}(t,{\bf k}) =
a^0_{\lambda,s}(t,{\bf k}) - i \alpha {{\bf e}_\lambda ({\bf k})\cdot {\bf d}(t) \over \sqrt{2\omega (2\pi)^3}}.
\end{align}
Exponential decay implies that the ${\bf d}(t)$-dependent second term does not contribute to the spectrum, which is therefore the same incorrect result $S_0(\omega)$ found when using the naive Coulomb gauge Hamiltonian $H_0(t)=H_0^0(t)$ \cite{milonni_natural_1989}. This calculation is easily extended to show more generally that the photonic source operator found using the EDA of the Hamiltonian $H_\alpha^{\alpha'}(t)$ in Eq.~(\ref{right}) is
\begin{align}
a^{\alpha'}_{\alpha,\lambda,s}(t,{\bf k}) =
a^{\alpha}_{\lambda,s}(t,{\bf k}) + i(\alpha-\alpha') {{\bf e}_\lambda ({\bf k})\cdot {\bf d}(t) \over \sqrt{2\omega (2\pi)^3}}.
\end{align}
Therefore, using any member of the equivalence class ${\mathscr S}_\alpha$ given by Eq.~(\ref{right}), results in the spectrum $S_\alpha(\omega)$.

\section*{Appendix B}

For completeness we here briefly address incorrect statements made in Refs.~\cite{settineri_gauge_2021,gustin_gauge-invariant_2023,di_stefano_resolution_2019} about time-dependent light-matter interactions. A more detailed discussion of all points below can be found elsewhere \cite{stokes_implications_2022}.\\

In Ref.~\cite{di_stefano_resolution_2019} it is claimed that ``{\em... during and after the switch off of the interaction, only the $\alpha=0$} (Coulomb) {\em gauge is well defined. Indeed, in the $\alpha\neq 0$ gauges the field momenta depend on the interaction strength"}.

On the contrary, every gauge that is defined for constant coupling strength (zero or otherwise) is defined during and after a time-dependent interaction. The fact that when no matter is present ${\bf E}=-{\bf \Pi}$ is transverse and unique, does not imply that in the presence of matter the field ${\bf E}_{\rm T}$, which equals $-{\bf \Pi}$ if $\alpha=0$, is independent of the light-matter interaction strength. Indeed, when matter is present ${\bf \Pi}$ possesses a coupling-dependent source part in {\em every} gauge \cite{stokes_implications_2022}.\\

In Ref.~\cite{settineri_gauge_2021} the Coulomb ($\alpha=0$) and Poincar\'e ($\alpha=1$) gauges are considered in the EDA. It is claimed that for introducing time-dependent interactions (notation adapted and bracketed text added for clarity) {\em ``we can consider $H_0$} [$\alpha=0$] {\em more fundamental than $H_1$} [$\alpha=1$]". 

One gauge cannot be considered more fundamental than another. This is a concept upon which the edifice of modern physics is built.\\

In Ref.~\cite{gustin_gauge-invariant_2023} it is claimed that through the prescription (\ref{wrong}) {\em ``... one can introduce time-dependent interactions... in either gauge without introducing any ambiguity or violating gauge invariance... this is done without treating any gauge as more fundamental than another"}.

On the contrary, the prescription (\ref{wrong}) defines the Coulomb gauge class ${\mathscr S}_0$ as universally correct, singling-out this gauge as preferred through a tacit assumption, namely, that time dependence arises solely through a specific gauge-fixing constraint. Ambiguity arises precisely because this assumption is not generally valid. In the case of a moving neutral material system, for example, the singling out of ${\mathscr S}_0$ results from the omission of the current associated with the gross motion, such that electric charge is not locally conserved. Since (by Noether) conservation of charge is implied by gauge symmetry (invariance), such a theory is not gauge-invariant.\\

It is further claimed in Ref. \cite{gustin_gauge-invariant_2023} that [bracketed text added] {\em ``In contrast to Ref. \cite{stokes_ultrastrong_2021}, which claims that introducing time-dependent light-matter interaction strengths necessarily breaks the gauge invariance of the fundamental light-matter Lagrangian, we note that this is circumvented by applying the time-dependent modulation of the interaction strength only to the transverse part of the interaction, since the transverse vector potential is gauge-invariant. This in fact produces equivalent results to the replacements made at the level of the Coulomb and multipolar gauge Hamiltonians described above,} [referring to Eq.~(\ref{wrong})] {\em and allows one to introduce gauge-invariant time-dependent couplings in a completely unambiguous way."}

If we interpret {\em ``transverse part of the interaction"} to mean interaction terms that involve only transverse fields, then this is the light-matter interaction (the interaction between the material and photonic quantum subsystems). Modulation of these terms is precisely what is considered in Ref.~\cite{stokes_ultrastrong_2021} (see also Ref.~\cite{stokes_implications_2022}). When considering a single {\em externally} bound charge, $-e$, as in Ref.~\cite{stokes_ultrastrong_2021}, this is achieved through the replacement $e \to e(t)$. The proof in Ref.~\cite{stokes_ultrastrong_2021} that this modulation results in non-equivalent Lagrangians when introduced in different gauges, is correct. We have now given a general presentation of the same mechanism by which this occurs in Sec.~\ref{general}, and a second example in Sec.~\ref{squid}.

If we instead suppose that {\em ``transverse part of the interaction"} means the interaction involving only the (gauge-invariant) transverse vector potential, then the claim of Ref.~\cite{gustin_gauge-invariant_2023} becomes again the incorrect claim that the Coulomb gauge interaction is preferred over any other. By definition of gauge-fixing, every fixed gauge possesses an interaction that can be written solely in terms of gauge-invariant observables, so no one gauge is distinguished by this property \cite{stokes_implications_2022}.

\bibliography{constraint.bib}

\end{document}